\documentclass[aps,showpacs,prep]{revtex4}
\usepackage{amssymb}
\usepackage{amsmath}
\usepackage{bm}

\begin{document}

\title{Quintessence interacting dark energy from induced matter theory of gravity}
\author{$^{1}$ L. M. Reyes \thanks{%
E-mail address: luzreyes@fisica.ufpb.br}, and  $^{1,2}$ Jos\'e Edgar Madriz Aguilar \thanks{%
E-mail address: madriz@mdp.edu.ar} }
\affiliation{$^{1}$ Departamento de F\'isica, DCI, Campus Le\'on, Universidad de Guanajuato, C.P. 37150, Le\'on Guanajuato, M\'exico, and \\
$^2$ Departamento de F\'isica, Facultad de Ciencias Exactas y Naturales, Universidad Nacional de Mar del Plata, Funes 3350,
C.P. 7600, Mar del Plata, Argentina. \\
E-mail: madriz@mdp.edu.ar}

\begin{abstract}
In the context of the induced matter theory of gravity, we investigate the possibility of deriving a 4D quintessential scenario where an interaction between dark energy and dark matter is allowed, and the dark energy component is modeled by a minimally coupled scalar field. Regarding the Ponce de Leon metric, we found that it is possible to obtain such scenario on which the energy densities of dark matter and dark energy, are both depending of the fifth extra coordinate. We obtain that the 4D induced scalar potential for the quintessence scalar field, has the same algebraic form to the one found by Zimdahl and Pavon in the context of usual 4D cosmology.
\end{abstract}

\maketitle

\vskip .5cm


\section{Introduction}

During the last years our understanding of the universe has changed. Just a few decades ago it was believed that the universe were passing for a period of a decelerated expansion. However, nowadays a great number of cosmologists agree that the universe is now in a period of accelerated expansion. This idea was originated in the supernova type Ia observations in 1998, which indicate that supernovaes are fainter than expected. Thus, at a fixed redshift, a specific supernovae must be more distant than it is expected in a decelerating universe \cite{Nova1,Nova2,Nova3,Nova4}. Faintness can be explained if it is regarded that the universe is currently passing for a period of accelerated expansion. The acceleration in the expansion is attributed to a component of the universe which exerts negative pressure: dark energy. The current observations suggest that this component has an equation of state parameter $\omega=-0.967^{+0.073}_{-0.072}$ \cite{OBS}. Dark energy is the largest contributor to the cosmological density, $\Omega_{de}=0.759\pm 0.034$, $\Omega_{mat}=0.241\pm 0.034$ \cite{OBS}. \\

Many cosmological scenarios have been proposed on which dark energy is modeled as a perfect fluid with an equation of state parameter ranging in the interval $-1<\omega<-1/3$. In this kind of models heading as quintessence models, the dark energy component is represented by a minimally coupled scalar field. Among the different quintessence models are the tracker models \cite{tracker}. These models it is considered an interaction between matter and dark energy, characterized by a constant ratio between its respective densities \cite{Inter}. These models appeared as an attempt to solve the cosmic coincidence problem.\\ 

Some other attempts to investigate the dark energy component and its cosmological implications have also been made (see for example \cite{Inter1,Inter2,Inter3,Inter4}). Other proposals include scalar fields with non-canonical kinetic terms, also known as k-essence models (see for example \cite{ke1,ke2,ke3,ke4}), and models based on modified theories of gravity \cite{MG1,MG2,MG3}. Theories in more than four dimensions are also been regarded to address the dark energy problem. Some of these theories are the membrane theory \cite{branes1,branes2,branes3,branes4} and the induced matter theory of gravity \cite{wbook,rept}. In membrane theory our ordinary 4D spacetime is viewed as a hypersurface (called the brane), embedded in a higher dimensional manifold (called the bulk). From the geometrical point of view the bulk is not necessarily Ricci-flat (in general $^{(5)}R_{ab}\neq 0$ ). Matter is confined to the brane by a variety of mechanisms, while gravity can freely propagate through the bulk.  \\

Parallel to the membrane theory there has been an intensive research in a proposal made by Wesson, Ponce de Leon and collaborators, known as the induced matter theory \cite{wbook,rept,Wesson}.  In this setting our universe is considered as a 4D hypersurface embedded in a 5D Ricci-flat ambient space. The extra coordinate is assumed non-compact and matter appears in 4D  as a consequence of the embedding. This matter is called induced matter and it is described by a conserved energy-momentum tensor of geometrical nature. Indeed, the induced matter depends of the extrinsic curvature of the embedded 4D hypersurface and of the extra component of the 5D metric. Thus a 5D spacetime, which is apparently empty, contains 4D sources of matter. The 5D vacuum field equations $^{(5)}R_{ab}=0$ specialized on the 4D hypersurface correspond to the 4D Einstein equations of general relativity, with a geometrical energy-momentum tensor describing classical properties of 4D physical sources of matter \cite{wbook,rept,Wesson}. This theory is mathematically supported by the well-known Campell-Magaard theorem that asserts that: {\it any analytic n-dimensional Riemannian space can be locally embedded in a $(n+1)$-dimensional Ricci-flat space}   \cite{CMT1,CMT2,CMT3,CMT4,CMT5}. In simple terms the curvature of the hypersurface appearing as a consequence of the embedding, is associated with the existence of sources of physical matter on 4D, as responsible by this curvature. This matter is modeled by a perfect fluid making thus possible to derive a geometrical equation of state for the induced matter, that of course will be dependent of the Ricci-flat 5D metric employed. Thus for every solution of $^{(5)}R_{ab}=0$, it as associated a geo\-me\-tri\-cal equation of state, that gives information about what kind of physical matter is being described geometrically.\\

Our goal in this letter is to investigate the possibility to derive in the context of the induced matter theory of gravity, a quintessential scenario with interacting dark energy modeled by a minimally coupled scalar field. The introduction of scalar fields within the framework of the induced matter theory has also been considered for inflationary scenarios for example in \cite{MBP1,MBP2,MBP3}. We found that such a scenario can be derived by using a solution of $^{(5)}R_{ab}=0$, known as the Ponce de Leon metric solution \cite{Ponce}. From this metric it is derived a geometrical equation of state that in certain limit can be associated to the equation of state of the present matter content of the universe. Then we modeled the induced matter by a perfect fluid where the energy density is composed by a component of dark matter ( that contains luminosity matter) and a component of dark energy. As we shall see, in this case both dark matter and dark energy are geometrically induced and thereby their corresponding energy densities are depending to the fifth coordinate. The scalar potential of the quintessence scalar field results to have an exponential form very similar to the one found by Zimdahl and Pavon in \cite{Inter}, in the context of the 4D standard cosmology. Thus, we organized the letter as follows. In section I we give a brief introduction. In section II we establish the basic equations of the induced matter theory. In section III we derive a quintessential scenario of interacting dark energy from a particular solution of the 5D field equations. Finally in section IV we give some final comments. Our conventions are: Latin indices run in the range $(0,1,...,4)$ with the exception of $i$ and $j$ that take values in the range $(1,2,3)$. Greek indices run from $(0,1,2,3)$. The metric signature we use is $(+,-,-,-,-)$, and we adopt units in which the speed of light $c=1$.

\section{The induced matter equations}

In the induced matter theory, we start by considering a 5D ambient space $({\cal M}^{5},g)$, endowed with a 5D metric $g_{ab}$, which is a solution of the 5D field equations $^{(5)}R_{ab}=0$. The 5D line element it is supposed having the form
\begin{equation}\label{a1}
dS_{5}^{2}=g_{ab}(x,l)dy^{a}dy^{b}=g_{\alpha\beta}(x,l)dx^{\alpha\beta}+\epsilon g_{ll}(x,l)dl^{2},
\end{equation}
with the parameter $\epsilon=\pm 1$ accounting for the signature and $l$ being the space-like extra coordinate assumed   non-compact. From the field equations $^{(5)}R_{ab}=0$ are constructed the conventional 4D Ricci tensor $^{(4)}R_{\alpha\beta}$ and the 4D Ricci scalar $^{(4)}R$. With these quantities, it is formed the 4D Einstein tensor $^{(4)}G_{\alpha\beta}=\,^{(4)}R_{\alpha\beta}-(1/2)\,^{(4)}Rg_{\alpha\beta}$, leaving the remaining terms in the component $^{(5)}R_{\alpha\beta}$, to construct an effective or induced energy-momentum tensor through the expression $^{(4)}G_{\alpha\beta}=8\pi GT_{\alpha\beta}^{(IM)}$, which reads \cite{Wesson}
\begin{equation}\label{a2}
8\pi G T_{\alpha\beta}^{(IM)}=\frac{\Phi_{,\alpha;\beta}}{\Phi}-\frac{\epsilon}{2\Phi^{2}}\left\lbrace\frac{\overset{\star}{\Phi}}{\Phi}\overset{\star}{g}_{\alpha\beta}-\overset{\star\star}{g}_{\alpha\beta}+g^{\lambda\mu}\overset{\star}{g}_{\alpha\lambda}\overset{\star}{g}_{\beta\mu}-\frac{1}{2}g^{\mu\nu}\overset{\star}{g}_{\mu\nu}\overset{\star}{g}_{\alpha\beta}+\frac{1}{4}g_{\alpha\beta}\left[\overset{\star}{g}^{\mu\nu}\overset{\star}{g}_{\mu\nu}+(g^{\mu\nu}\overset{\star}{g}_{\mu\nu})^{2}\right]\right\rbrace,
\end{equation}
where $\Phi^{2}=g_{ll}$, the coma is denoting partial derivative, the semicolon denotes 4D covariant derivative and the star denotes partial derivative with respect to the extra coordinate $l$.\\
 
To our purposes, let us to consider a 5D line element of the form 
\begin{equation}\label{a3}
dS_{5}^{2}=g_{ab}dx^{a}dx^{b}=e^{2A(\tau,l)}d\tau^{2}-e^{2B(\tau,l)}\delta{ij}dx^{i}dx^{j}-e^{2C(\tau,l)}dl^2,
\end{equation}
where $\tau$ is a time-like coordinate and $x^{i}$ are space-like coordinates. The functions $A(\tau,l)$, $B(\tau,l)$ and $C(\tau,l)$ are well behaved metric functions.  The 5D field equations $^{(5)}R_{ab}=0$ or equivalently $^{(5)}G_{ab}=0$ for the metric  (\ref{a3}) can be written as
\begin{eqnarray}\label{a4}
&&-3(\dot{B}^{2}+\dot{B}\dot{C})e^{-2A}+3(\overset{\star\star}{B}+2\overset{\star}{B}^{2}-\overset{\star}{C}\overset{\star}{B})e^{-2C}=0,\\
\label{a5}
&& 3(\overset{\star\, .}{B}+\dot{B}\overset{\star}{B}-\dot{B}\overset{\star}{A}-\overset{\star}{B}\dot{C})e^{-2A}=0,\\
\label{a6}
&&-(2\ddot{B}+3\dot{B}^{2}+\ddot{C}+\dot{C}^{2}+2\dot{B}\dot{C}-a\dot{A}\dot{B}-\dot{A}\dot{C})e^{-2A}+(2\overset{\star\star}{B}+3\overset{\star}{B}^{2}+\overset{\star\star}{A}+\overset{\star}{A}^{2}+2\overset{\star}{B}\overset{\star}{A}-2\overset{\star}{C}\overset{\star}{B}-\overset{\star}{A}\overset{\star}{C})e^{-2C}=0,\\
\label{a7}
&&-3(\ddot{B}+2\dot{B}^{2}-\dot{A}\dot{B})e^{-2A}+3(\overset{\star}{B}^{2}+\overset{\star}{B}\overset{\star}{A})e^{-2C}=0,
\end{eqnarray}
where the dot is denoting derivative with respect to the time like coordinate $\tau$. As it is usually done in the induced matter approach, by means of the formal identification of the geometrical energy-momentum tensor (\ref{a2}) with an energy-momentum tensor of a perfect fluid $T_{\alpha\beta}=(\rho+p)u_{\alpha}u_{\beta}-pg_{\alpha\beta}$, we obtain on the geometrical background (\ref{a3}), an energy density $\rho$ and a pressure $p$ given respectively by the expressions 
\begin{eqnarray}\label{a8}
8\pi G \rho & = & (\ddot{C}+\dot{A}\dot{C})e^{-2A}+(\overset{\star}{C}\overset{\star}{A}-\overset{\star\star}{A}-\overset{\star}{A}^{2}+3\overset{\star}{B}^{2})e^{-2C},\\
\label{a9}
8\pi G p & = & -[\dot{B}\dot{C}e^{-2A}+(\overset{\star}{C}\overset{\star}{B}-\overset{\star\star}{B}+2\overset{\star}{A}\overset{\star}{B})e^{-2C}].
\end{eqnarray}
Employing the equations (\ref{a8}) and (\ref{a9}), we can define a geometrical equation of state in the form
\begin{equation}\label{a10}
\omega_{g}\equiv\frac{p}{\rho}= -\left[1-\frac{(\ddot{C}+\dot{C}^{2}-\dot{B}\dot{C})e^{-2A}+(\overset{\star\star}{B}-\overset{\star}{C}\overset{\star}{B}-2\overset{\star}{A}\overset{\star}{B}-\overset{\star\star}{A}-\overset{\star}{A}^{2}+3\overset{\star}{B}^{2})e^{-2C}}{(\ddot{C}+\dot{C}^{2})e^{-2A}-(\overset{\star\star}{A}+\overset{\star}{A}^{2}-3\overset{\star}{B}^{2})e^{-2C}}\right].
\end{equation}
One of the main ideas in the induced matter theory, is that the 4D Riemann curvature that appears on the 4D spacetime embedded in the 5D Ricci-flat spacetime, is interpreted as generated by the presence of sources of physical matter. In the particular case of a 5D ambient space with the metric (\ref{a3}), the physical matter causing the 4D curvature must satisfy an equation of state given by (\ref{a10}). This is like if we instead of seeing directly to the physical matter, we look just the 4D curvature that it generates, and then via its geometrical equation of state we determine the nature of that matter. For example, if the equation (\ref{a10}) for a specific value of the metric functions $A(\tau,l)$, $B(\tau,l)$ and $C(\tau,l)$ were reduced to $p=1/3\rho$, then we could say that the curvature on 4D is generated by radiation. This perspective seems to be a suitable alternative for addressing some cosmological issues and in particular, to model dark energy. Many other solutions in this context has been extensively studied by P. Wesson, J. Ponce de Leon and collaborators. Some of them can be seen for example in \cite{wbook,rept,Wesson}.

\section{A quintessence scenario with interacting dark energy}

With the idea to derive a quintessential scenario in the context of the induced matter theory, let us to regard a particular case of the line element (\ref{a3}), given by 
\begin{equation}\label{b1}
dS_{5}^{2}=l^{2}d\tau^{2}-a^{2}(\tau)l^{[2\sigma/(\sigma-1)]}\delta{ij}dx^{i}dx^{j}-(\sigma -1)^{-2}\tau^{2}dl^{2},
\end{equation} 
where $a(\tau)=\tau ^{\sigma}$, $\delta_{ij}$ is the Kronecker delta, $\sigma$ is a constant, the $x^{i}$ are the 3D cartesian coordinates and $l$ in this case is dimensionless. This metric is a solution of the 5D field equations $^{(5)}R_{ab}=0$, and has been first obtained by J. Ponce de Leon in \cite{Ponce}. The 5-velocities $U^{a}\equiv dy^{a}/dS_{5}$ of the 5D observers satisfy the system 
\begin{eqnarray}\label{b2}
\frac{dU^{a}}{dS_{5}}+\,^{(5)}\Gamma^{a}_{bc}U^{b}U^{c} &= & 0,\\
\label{b3}
g_{ab}U^{a}U^{b} &=& 1.
\end{eqnarray}
In order to get a cosmological description, let us to choose comoving observers with respect to the 3D space by taking $U^{i}=0$ i.e.  $U^{a}=(U^{\tau},0,0,0,U^{l})$. The components $U^{\tau}$ and $U^{l}$ according to the equations (\ref{b1}),  (\ref{b2}) and (\ref{b3}) have the form
\begin{equation}\label{b4}
U^{\tau}=\mp \frac{1}{\sqrt{\sigma(2-\sigma)}}\frac{1}{l},\qquad U^{l}=\pm\frac{(\sigma -1)^2}{\sqrt{\sigma(2-\sigma)}}\frac{1}{\tau}.
\end{equation}
Now we shall assume that the 5D spacetime can be foliated by a family of 4D hypersurfaces $\Sigma$ defined by the expression $l=constant$. Then the metric on every hypersurface, say $l=l_0$, will be the induced metric, given in this case by 
\begin{equation}\label{b5}
dS_{4}^{2}=l_{0}^{2}d\tau^{2}-a^{2}(\tau)l_{0}^{[2\sigma/(\sigma-1)]}\delta{ij}dx^{i}dx^{j}.
\end{equation}
However, according to (\ref{b4}) the hypersurface $\Sigma$ is not a confining one. So in order to get $\Sigma$ a confining hypersurface, we implement the coordinate transformation $t=l_{0}\tau$, leaving  the rest of the coordinates unaltered. In the new coordinates we can see that $g_{tt,l}=0$, so it can be easily shown that the system (\ref{b2})-(\ref{b3}), for a class of observers satisfying $U^{i}=0$ and $U^{l}=0$, leaves to the usual 4D geodesic equation. Hence, the class of observers living on every confining hypersurface $\Sigma$ is  $U^{a}=(U^{t},0,0,0,0)$ and correspond to 4D comoving observers. The line element (\ref{b5}) therefore becomes
\begin{equation}\label{b6}
dS_{4}^{2}=dt^{2}-a^{2}_{eff}(t)\delta{ij}dx^{i}dx^{j},
\end{equation}
where $a_{eff}(t)=t^{\sigma}l_{0}^{\sigma(2-\sigma)/(\sigma-1)}$ and $t$ is the cosmic time. This metric corresponds to a FRW metric 3D spatially flat.  \\
Inserting the metric (\ref{b1}) in the expression (\ref{a10}), we obtain an equation of state for the induced matter in 4D given by 
\begin{equation}\label{b7}
p=-\left(1-\frac{2}{3\sigma}\right)\rho,
\end{equation}
which is a geometrical equation of state studied for example in \cite{wbook}. One remarkable aspect of this equation is that for $|2/(3\sigma)|\ll 1$ (or zero), this equation describes practically a vacuum equation of state. As it is well known an equation of state of this nature appears in the description of periods of accelerated expansion of the universe: inflation and the present epoch.  Thus the geometrical equation of state (\ref{b7}) seems to be suitable to describe the matter content of the present period of accelerated expansion.\\
The Friedmann equations on $\Sigma$ according to (\ref{b6}), then read
\begin{eqnarray}\label{b8}
\frac{3}{a_{eff}^2}\left(\frac{d}{dt}a_{eff}\right)^{2}&=&8\pi G \rho,\\
\label{b9}
\frac{2}{a_{eff}}\frac{d^{2}}{dt^2}a_{eff}-\frac{1}{a_{eff}^2}\left(\frac{d}{dt}a_{eff}\right)^{2}&=&-8\pi G p,
\end{eqnarray}
where $\rho=3\sigma^{2}/t^{2}$ and $p=\sigma(2-3\sigma)/t^{2}$. From the equations (\ref{b8}) and (\ref{b9}), it can be easily seen that the geometrical quantities $\rho$ and $p$ are geometrically mimicking the total energy density and the total pressure of the universe. Thus, it seems appropriated to split $\rho$ and $p$ as follows 
\begin{equation}\label{b10}
\rho=\rho_{m}+\rho_{\phi},\qquad p=p_{\phi},
\end{equation}   
being $\rho_{m}(t)$ the density of dark matter (including the visible matter), $\rho_{\phi}$ the density of dark energy and $p_{\phi}$ the pressure of dark energy. As we are interested in modeling the dark energy by a quintessence model, we will assume that the dark energy can be modeled by a 4D scalar field $\phi(t)$. Then $\rho_{\phi}=(1/2)(d\phi/dt)^{2}+V(\phi)$ and $p_{\phi}=(1/2)(d\phi/dt)^{2}-V(\phi)$, where $V(\phi)$ is the scalar potential which must be determined.\\ 

If we assume that dark energy do not evolve independently of dark matter, the conservation of the energy-momentum tensor of the induced matter (\ref{a2}) results in the expressions
\begin{eqnarray}\label{b11}
\frac{d\rho_{m}}{dt}+\frac{3}{a_{eff}}\frac{da_{eff}}{dt}\rho_{m}=f(t),\\
\label{b12}
\frac{d\rho_{\phi}}{dt}+\frac{3}{a_{eff}}\frac{da_{eff}}{dt}(\rho_{\phi}+p_{\phi})=-f(t),
\end{eqnarray}
where f(t) is an interaction function such that when $f(t)=0$, it means that dark energy and dark matter do not interact. Assuming that the equation of state for dark energy has the form $p_{\phi}=\omega_{\phi}\rho_{\phi}$, the equations (\ref{b11}) and (\ref{b12}) yield
\begin{equation}\label{b13}
\frac{d}{dt}[(1+r)\rho_{\phi}]+\frac{3}{a_{eff}}\frac{da_{eff}}{dt}\left(1+\frac{\omega_{\phi}}{1+r}\right)(1+r)\rho_{\phi}=0,
\end{equation}
where  $r(t)=\rho_{m}/\rho_{\phi}$ being the ratio between matter density and dark energy density, which in general is a function of time. The equation (\ref{b12}) has for solution 
\begin{equation}\label{b14}
\rho_{\phi}=\frac{\rho_{\phi_0}}{1+r}e^{-3\int H_{eff}\left(1+\frac{\omega_{\phi}}{1+r}\right)}dt\, ,
\end{equation}
where $H_{eff}(t)=(1/a_{eff})[da_{eff}/dt]$ is the effective Hubble parameter. By employing the equations (\ref{b10}) and (\ref{b14}) we obtain the relation 
\begin{equation}\label{b15}
r(t)=\frac{2+3\sigma(1+\omega_{\phi}(t))}{3\sigma-2}.
\end{equation}
In the particular case when $r=constant$, the expression (\ref{b14}) can be written as
\begin{equation}\label{b16}
\rho_{\phi}=\rho_{\phi_0}\left(\frac{a_{eff}}{a_0}\right)^{-\gamma}=\rho_{\phi_0}\,l_{0}^{-\,\frac{\gamma\sigma(2-\sigma)}{\sigma-1}}t^{-\gamma\sigma},
\end{equation} 
being $\gamma=3[1+(\omega_{\phi}/(1+r))]$ and where we have taken the present value of the effective scale factor to be  $a_{0}=1$. From the expressions (\ref{b12}) and (\ref{b14}) the interaction function $f(t)$ read
\begin{equation}\label{b17}
f(t)=\frac{\frac{dr}{dt}-3r\omega_{\phi}H_{eff}}{1+r}\,\rho_{\phi},
\end{equation}
which for the case $r=constant$ clearly reduces to 
\begin{equation}\label{b18}
f(t)=-\frac{3r\omega_{\phi}H_{eff}}{1+r}\,\rho_{\phi}.
\end{equation}
Now using the geometrical equation of state (\ref{b7}) and the equation (\ref{b10}) we obtain that 
\begin{equation}\label{b19}
\frac{\omega_{\phi}}{1+r}=-\left(1-\frac{2}{3\sigma}\right).
\end{equation}
This relation in terms of $\gamma$ gives the simple relation $\gamma=2/\sigma$. Observations of CMBR and of supernovae Ia suggest that $r\simeq 1/3$ \cite{OBS}. Thus for $-1\leq\omega_{\phi}\leq -0.894$ the expression (\ref{b19}) restricts $\sigma$ to range in the interval $2.02327\leq\sigma\leq 2.6666$. From the equations (\ref{b16}),  (\ref{b8}) and (\ref{b10}), it can be easily shown that the density of matter $\rho_{m}$ evolves as
\begin{equation}\label{b20}
\rho_{m}=\rho_{m_0}\left(\frac{a_{eff}}{a_{0}}\right)^{-\gamma}=\left(3\sigma^{2}-8\pi G\rho_{\phi_0}l_{0}^{-\frac{2(2-\sigma)}{\sigma-1}}\right)t^{-2},
\end{equation}
where $\rho_{m_0}=[(3\sigma^{2})/(8\pi G)]l_{0}^{2(2-\sigma)/(\sigma -1)}-\rho_{\phi_0}$. Making use of (\ref{b7}), the system (\ref{b10}) can be rewritten as 
\begin{eqnarray}\label{b21}
\frac{3\sigma^2}{t^2}&=&\frac{1}{2}\left(\frac{d\phi}{dt}\right)^{2}+V+\rho_{m},\\
\label{b22}
\frac{\sigma(2-3\sigma)}{t^{2}}&=&\frac{1}{2}\left(\frac{d\phi}{dt}\right)^{2}-V.
\end{eqnarray}
The scalar field of quintessence $\phi(t)$ from the equations (\ref{b21}) and (\ref{b22}) is found to be
\begin{equation}\label{b23}
\phi(t)=\phi_{0}+\lambda_{0}\ln\left(t/t_{0}\right),
\end{equation}
where $\lambda _{0}=[\sigma(2-3\sigma)]+8\pi G\rho_{\phi_0}l_{0}^{-2(2-\sigma)/(\sigma-1)}]^{1/2}$. Similarly, from the expressions (\ref{b21}) and (\ref{b22}) we find that the scalar potential as a function of the time is given by
\begin{equation}\label{b24}
V=\frac{1}{2}\left[8\pi G\rho_{\phi_0}l_{0}^{-\frac{2(2\sigma)}{\sigma-1}}-\sigma(2-3\sigma)\right]t^{-2}.
\end{equation}
Finally, regarding the equations (\ref{b23}) and (\ref{b24}), the potential $V$ as a function of the field of quintessence $\phi$ results of the form
\begin{equation}\label{b25}
V(\phi)=V_{0}e^{-\frac{2}{\lambda_0}(\phi-\phi_0)},
\end{equation}
where $V_{0}=(1/2)[8\pi G\rho_{\phi_0}-\sigma(2-3\sigma)l_{0}^{2(2-\sigma)/(\sigma-1)}]$ denotes the present value of the scalar potential $V_{0}=V(\phi_0)$. This shows that a quintessential scenario with late-time constant ratio between matter and dark energy can be realized within the framework of the induced matter theory of gravity. An exponential potential of the form of (\ref{b25}) has been obtained by Zimdahl and Pavon in \cite{Inter} within 4D quintessential scenarios with interacting dark energy. 

\section{Final comments}

In this letter, we have shown that it is possible to obtain a quintessential scenario of interacting dark energy, in the context of the induced matter theory of gravity. Using a particular metric solution of the 5D field equations known as the Ponce de Leon metric, we choose a class of observers and coordinates, such that the 4D hypersurface $\Sigma:l=l_{0}=constant$, is a confining hypersurface. Making use of the energy-momentum tensor of induced matter, we derive the geometrical equation of state (\ref{b7}) on every hypersurface $\Sigma$. This equation of state has the characteristic that when $|2/(3\sigma)|\ll 1$ (or zero), it is a good candidate to describe the matter content of the present universe. Next we give a physical identity to the induced matter via a formal identification of its energy density $\rho$ and pressure $p$, with the total energy density and the total pressure of the present matter content of the universe, at cosmological scales, as it is indicated in the expression (\ref{b10}).\\
Representing the dark energy component by a scalar field $\phi(t)$ and assuming that dark energy do not evolve independently of dark matter, we found general expressions for the energy density of dark matter $\rho_{\phi}$ and the interaction function $f(t)$, under the assumption that the ratio between matter and dark energy is a function of the time $r(t)=\rho_{m}/\rho_{\phi}$. In particular when this ratio is constant, we obtain that the scalar field of quintessence has a  logarithmic dependence of the time, as shown in the equation (\ref{b23}). Finally, it is obtained that the corresponding scalar potential $V(\phi)$ has essentially the same algebraic form that the one obtained by Zimdahl and Pavon in \cite{Inter}. The difference relays in the fact that here both $V_{0}$ and $\lambda_{0}$ are depending of the extra coordinate $l=l_0$. The exact expression for this potential is shown in the equation (\ref{b25}).

\section*{Acknowledgements}

\noindent L. M. R. and J.E.M.A acknowledge CONACYT M\'exico  for financial support.

\end{document}